Title:              C. RAGOONATHA CHARY AND HIS VARIABLE STARS

Authors:            N. Kameswara Rao , A.Vagiswari, Priya Thakur & Christina Birdie

E-mail:             nkrao@iiap.res.in, vagiiap@iiap.res.in, priya@iiap.res.in, chris@iiap.res.in

Affiliation:        Indian Institute of Astrophysics, Koramangala, Bangalore 560 034

ABSTRACT

C. Ragoonatha Chary, the first assistant at Madras Observatory during 1864 to 1880 was not only a celebrated observational astronomer but also a person who emphasized the need for incorporating modern observations based improvements into the traditional methods of astronomical calculations. He was one of the first few people who argued for establishment of independent modern Indian observatory for education and training. He was credited with the discovery of two variable stars R Reticuli and another star whose identity is uncertain. The person and his variable star discoveries are discussed.

## 1    INTRODUCTION

The phenomenon of stellar variability is of great importance in very wide areas of astrophysics including stellar structure, stellar evolution, distance scales (through period- luminosity relations), dust formation etc. The study of variable stars in a systematic way had only started by mid nineteenth century. There were only 18 variable stars known in 1844 (Hogg 1984). The status of the variable star research around that time has been well summarized by Clarke (1903), Hogg (1984), Orchiston (2000) etc. It was only practiced by few enthusiasts like Pogson, Baxendalls etc particularly encouraged by people like Arglanders and Pickring. However the fascination to study the light variability of stars was always not been thought appropriate for government (British) supported observatories like Madras Observatory, where the priority and main purpose being to produce catalogues of stellar positions from meridian observations. In spite of it Norman Robert Pogson, soon after arriving in Madras in 1860 as the Government astronomer introduced the observations of variable stars as part of the programme and inspired people like Ragoonatha Chary to enthusiastically pursue this.

Chinthamani Ragoonatha Chary was the first Indian who communicated (published) a paper to the Royal Astronomical Society (RAS) as well as the first Indian to become Fellow RAS. His name was proposed by N.R. Pogson, the then Government Astronomer at Madras Observatory and E.B. Powell, the then Director General of public instruction. He was elected as a FRAS on 12 January 1872. Ragoonatha Chary was not only a keen observer, who devoted almost his whole life to observational astronomy at Madras Observatory but also , more importantly, he was a passionate promoter of modern astronomy (and science) and a reformer who endeavored to integrate modern observational results and physical phenomenon into the classical Sidhanthic Astronomy usually practiced then. His efforts in formulating and bringing out the 'Drigganita Panchang' (an Almanac-calculated, based on observations), for several years is well described by Venkateswaran (2009). While commenting on his own efforts in writing a treatise titled 'Jyothisha Chinthamany' that was to contain rules, formulae and tables based on the 'English' methods of calculation for the guidance of our Sidhantis, he champions the cause of observational astronomy by urging people to raise funds "to establish an observatory, to serve as a school for the instruction of Hindu students desirous to qualify in practical astronomy". He emphasizes further

> I earnestly commend this movement (raising funds for an observatory as well as to help publish his treatise) to all native noblemen and wealthy gentlemen in the Presidency(Madras), as well as throughout India, who are interested in the improvement of their fellow countrymen, and beg them to join heartily in a design which aims at promoting a most fascinating branch of knowledge, the cultivation of which although under besetting difficulties and imperfections, is now and always has been highly prized by Hindus throughout the country.

He acquired the skills of a great observer by hard work and devotion. He joined the Madras Observatory as a young 'cooli' (according to Venkateswaran) at an age of 18 years and rose to the position of first (and head) assistant to the Astronomer. He not only did social duties as an observer and calculator at the observatory, but also used to continue his observations even at his residence in his off duty time with his own instruments (he even offered them for use at the proposed Indian observatory whenever it got ready).  In the process he discovered two (?) variable stars, R Reticuli and V(?) or U(?) Cephei.

## 2    AS A PRIVATE PERSON

Not much is known about his non-professional life. Ragoonatha Chary (the name has also been spelt by some as Raghunatha Chari) must have had a good Sanskrit education in his early childhood based on his proficiency at interpreting and analyzing the Sidhanthic astronomy texts. He apparently hailed from a family of panchang (almanac) makers according to Venkateswaran (2009). None of the available records mention his date of birth. Pieced together from various remarks by Pogson about his length of service, the age of joining the service etc. we infer the probable year of birth as 1822. However the obituary notices of RAS Fellows and Associates mention it as 1828? He was a self taught man and managed to acquire mathematical knowledge so as to be able to predict occultation of stars in the Sun's path during the total solar eclipse of 1868 and much more (Ragoonatha Chary, 1868). According to Pogson` he possessed sufficient skill and energy to make additional observations, worthy of the reputation of the Observatory and beneficial to science' (Pogson: 1861a). Ragoonatha Chary lived in Nungumbakam village (Madras Almanac: 1867) and close to the observatory. Pogson remarked in a footnote to his log sheets listing the observations of occultation of Venus on 3 November 1872 that Ragoonatha Chary observed the event 'from his private residence about five eighths of a mile distant nearly due south of the observatory'. Limited information is available about his family. He mentions that his son Chinthamani Raghava Chary assisted him in preparation of Drigganita Panchang for the year 1880. He also seems to have inducted one of his relatives P. Raghavachari as a third assistant at Madras Observatory during 1877. He seems to have participated in socially beneficial activities. He was the executive committee president of Madras Hindu Janopakara Nithee (general benefit fund) established at Pursewakum on 1 February 1861. The management of the fund was entrusted to a committee of 16 members. Along with his colleague T. Mootoosowmy Pillay, Ragoonatha Chary served as an executive committee trustee of Madras Hindu Draviya Sakara Nidhi (Saving fund) that was established in Nungumbakam on 1 July 1861. He edited for twelve years the astronomical portion of Asylum Press Almanac. He was also known to give public talks on various astronomical (and science) topics.

## 3      AT MADRAS OBSERVATORY

Ragoonatha Chary had a long career of about forty years at Madras Observatory. He was recruited by T.G. Taylor probably around 1840 and worked with successive Directors of the observatory W.S. Jacob, who communicated his first paper to Monthly Notices of RAS, W.K. Worster, J.F. Tennant and N.R. Pogson who were all impressed by his astronomical skills. His main mentor however had been Pogson. Soon after his arrival at Madras, Pogson announced the discovery of the minor planet 'Asia', the first from this part of the world (Pogson,1861 b), in which he compliments Ragoonatha Chary's participation "the second observation, on 20 April, was taken and reduced by my fourth native assistant, Ragoonatha Chary, who readily comprehends and most willingly executes whatever I may recommend to his notice." Ragoonatha Chary's zeal to communicate the modern developments in astronomy to general public and to make them aware of the physical nature of the celestial events so as to dispel the blind superstitions has been encouraged by Pogson. The two pamphlets Ragoonatha Chary prepared for the benefit of general public, on the occasion of the two total solar eclipses on 18 August 1868 and 12 December 1871 that passed through South India received considerable support. The one describing 1871 eclipse which was brought out in four regional languages, was strongly endorsed and recommended by Pogson to the Chief Secretary of the Madras Government for financial help

> consider it well calculated to instruct and at the same time to dispel the superstitious fears of the ignorant native masses, while it goes further and by its quotations from the puranahs, and other weighty arguments show that the absurd notions which render eclipses so alarming are not authorized by their own religious writings. He also shows by comparison of results, the vast inferiority and great inaccuracy of the calculations of Hindoo astronomers and urges them to abandon their worthless methods and antiquated tables and avail themselves of modern improvements. (Pogson's letter of 16 August 1871 to R.S. Ellis).

During the historic total solar eclipse of 18 August 1868, where the element Helium was discovered, Ragoonatha Chary was given the responsibility of conducting the observations of the eclipse from Vunparthy, a village located at forty-v e miles north of Kurnool. He unfortunately had a cloudy weather and could only comment on the amount of darkness during the eclipse. He was one of the main persons chosen for the Madras Observatory's expedition team to conduct observations during the total solar eclipse of 1871 at Avenashy, in the Coimbatore District. His role in the 12 December 1871 total eclipse was to conduct "general observations, consisting of careful micrometrical measurements of the cusps, of conspicuous prominences, of the extant and figure of corona, amount of darkness, visibility of stars, and accurate times of the various phenomena, may be chiefly entrusted to C. Ragoonatha Chary, the first Native assistant" (Pogson's letter of 3 July 1871 to the Acting Chief Secretary to Madras Government, Fort St. George). He continued similar work of Coronal observations at 6 June 1872 annular solar eclipse that passed through Madras, with Lerebours equitorial (Pogson, 1872).

However Ragoonatha Chary's main routine role at Madras Observatory was to obtain with Meridian Circle, observations of stars and their reductions for preparing the Madras Catalogues, which he continued till 1878 as long as his deteriorating health permitted him. A new transit circle, that was originally designed by G.B. Airy for Royal Observatory Greenwich was duplicated and sent to Madras (Sen, 1989). It was installed and put to regular use by Pogson since 1862. This was the main instrument Ragoonatha Chary worked with. Till the end of 1878 a total number of 35,681 observations were collected with that instrument out of which a fair number of them were by Ragoonatha Chary.

## 4  VARIABLE STARS

Soon after N.R. Pogson joined Madras Observatory as Government Astronomer in 1861 he included observations of variable stars and minor planets into the main programmes of the observatory. Pogson was a pioneer in the observations of variable stars having discovered several of them both at Oxford as well as at Madras (a total of 21 of them). He was also in the process of preparing The Variable Star Atlas (Pogson, Brook, Turner, 1908). He used to provide ephemeris of variables annually in Monthly Notices of RAS. In 1861 the total listed known long period variableswere55 (!) in 1862 they rose to 60. He included several known variable stars into the regular observational schemes at Madras Observatory, even with the meridian circle. For example the well known irregular variable R *Coranae Borealis* was observed on different times both at light maximum as well as below maximum light between 20 May 1863 and 4th July showing brightness variation from 6.1 to 9.0. Ragoonatha Chary himself contributed four of them on three occasions (the others were by Mootoosawmy Pillay). The magnitude scale adopted was earlier devised by Pogson (Pogson, Brook, Turner, 1908). Ragoonatha Chary's observations (positions) were utilized by Webbink (1978) in re-identifying the recurrent novae U Sco that was discovered as a variable by Pogson from Madras on 20 May 1863. Due to this new and exciting area of activity undertaken by Madras Observatory Ragoonatha Chary became a well versed variable star observer. It was only expected that he would soon discover a new variable star himself.

### 4.1  R Reticuli

It did happen on the night of January 1866. Pogson (1867) describes this event in the Annual report of the Madras Observatory thus 'The detection of a new and interesting variable star, far south, is due to the First Native Assistant C. Ragoonatha Chary.' It was first casually observed with the Meridian Circle on 9 February 1864, as an ordinary star of 8 1/2 magnitude (by Moottooswamy), but when next looked for, in January 1866, was no longer visible in the dark field of the circle telescope, 5 1/2 inches in aperture. It must then have been under the 12th magnitude. It was, however, re-observed on 16 January 1867, and its variability there by proved by the above named observer. The subsequent light comparisons, on twenty six nights up to 7 April showed that it attained a maximum brilliancy of 7 3/4 - magnitude about the middle of February and that its period of variation or interval between two successive maxima is most probably about nine months. The place of this new variable star, which will be known as R. Reticuli Var.1, is Right Ascension $4^h\ 32^m\ 6.1^s$ and North Polar distance $153°\ 19'14"$ for the epoch of 1 January 1860. Figure 3 shows the observations of R Reticuli in 1867.

It has been further followed at Madras Observatory (Pogson, Brook, Turner, 1908) as well as at Harvard Observatory (Campbell, 1926) etc. Now, it is a well known large amplitude Mira type variable star with a period of 278.32 days, (B-V) 0 of 1.34 and E (B-V) =0.11. The spectral type varies between M4e to M7.5e. Allen *et al* (1989) have discovered it as a SiO maser source 86 GHz. Fairly a large body of literature exists about this star now.

### 4.2  The Mysterious second variable star discovery (?)

The obituary of Ragoonatha Chary that appeared in Madras Mail of 7th February 1880 mentions that Ragoonatha Chary has also discovered another variable star, V Cephei, in 1878. The same discovery of V Cephei was mentioned again in the obituary of Ragoonatha Chary that appeared in Monthly Notices of Royal Astronomical Society (MNRAS: 41, Feb 1881:180) but the source is attributed to Madras Mail. However the Madras Almanac, of 1880 lists the star that was discovered as a variable, as U Cephei. Surprisingly no mention of this discovery of a new variable star (V or U Cephei) appears in the annual reports of Madras Observatory for the years 1878, 1879 by Pogson (who was a great champion of variable stars and took great pleasure in writing about R Reticuli) nor does it appear in the Pogson's Atlas of Variable stars.

However, the discovery of the variability of V Cephei (V = 6.59, B-V =0.05, U-B =0.05, sp. type A3V) was credited to C.S. Chandler in 1882 (Hoffleit, 1985) who found a variation of 0.7 magnitudes and assumed that the star to be either a long period or irregular variable. Hoffleit (1985) discusses the nature of variability of this object in detail as some early observers including W.J. Luyten, found the star to be a variable of about 0.5 magnitude whereas neither E.C. Pickering nor H. Shapley could find variations in light. The spectral type of A3V does not support the existence of pulsation either. Milton, Williams, & Hoffleit (1988) further discuss the nature of the star from additional photoelectric photometry and conclude that the star has not varied by more than 0.02 magnitude in a period of 428 day interval and most likely the star is not a variable at all.

The variability of U Cephei in the literature has been credited to Ceraski (1880) who discovered the variation on 23 June 1880 i.e. - after the death of Ragoonatha Chary. It is a well known Algol type binary with a period of 2.49 day period. Soon after the discovery in 1880, Baxandells, both Senior and Junior, who were brother-in-law and nephew (respectively) of Pogson, observed this star. It is unlikely that they would be unaware of any discovery of this star had it been made in Madras Observatory (early studies on the variability see Yendell, 1903; Shapley, 1916). Ragoonatha Chary was not mentioned any wherein these discoveries.

So what is it then that was discovered by Ragoonatha Chary?

Searching the published records of Madras Observatory's "Separate Results of Madras Meridian Circle Observations in 1878" (Pogson1887) shows an entry, "U Cephei, var 5" (see Figure 4) which lists the times of observations of the star, the magnitude, the mean positions for 1878 and the observer. It was observed on v e occasions by three different observers Raghavachari (R), Ragoonatha Chary (C.R) and Mootoosawmy Pillay (M). The position of the object as listed by all the observers is the same, although the magnitude varies from 5.0 to 9.0. The fainter magnitudes were observed by Ragoonatha Chary i.e. 8.2 and 9.0 from its brightness maximum of magnitude 5.0. Again the same object was observed in 1880 (Figure 5) and listed in `Separate Results of Madras Meridian Circle Observations in 1880' (Pogson and Smith 1893) as U Cephei, Var. 5. It was observed on five occasions by Mootoosawmy Pillay showing a variation in brightness of 6.5 to 7.0 magnitude (1.5to 2 magnitudes fainter than the brightest that was observed in 1878). All these catalogues were published in 1893 after the death of Pogson under the supervision of Michie Smith. Usually var. 5 refers to the 5th variable discovered in the constellation. In the notation (Arglander's) practiced by variable star observers it should refer to V Cephei than to U Cephei (which would be 4th variable). More importantly the coordinates given in these catalogues do not match with that of either U Cephei (discovered by Ceraski) or V Cephei (discovered by Chandler). The star referred by Madras Catalogues is more than 2 hours west and 13˚ south of V Cephei and more than 3 hours west and more than 11.5˚ south of U Cephei. Then which is the star mentioned by Madras Catalogues?

Search of 10 arc minute field around the expected position of the star showed agreement of position with HR 8342 (=HD 207636). We calculated the mean position of this star for epoch of 1878 using the proper motion given by Simbad (Hipparcos) as the following: (α, δ): $21^h 44^m 51.5^s + 69˚ 35'7.2"$.

The position of the Madras Variable (figure 4) is: (α, δ): $21^h 44^m 51.4^s + 69˚ 35'8.1"$.

The coordinates match very well indeed. So the Madras variable could very likely be identified with HR 8342.

HR 8342 is not known to be a variable both in light as well as radial velocity. The three measurements listed in Simbad on different occasions show a mean value of $-2 \pm 1$ km s$^{-1}$. The spectral type is A0 V. The Hipparcus parallax gives a distance of $151 \pm 10$ parsecs and with a V = 6.45, the Mv obtained is 0.56, which is consistent with the spectral type of A0 V (Allen, 1973). The A0 V stars are not known to be variable, that too by four magnitudes. The possibility that it could be a binary also seems unlikely. For example the Algol system U Cephei has its primary as B8 V and similar magnitude, V = 6.92. It shows light variation ranging from 6.9 to 9.2 magnitude. However the infrared colors clearly show an excess (for a B8 V star) suggesting the presence of a companion. The colors of U Cep are: B-V = 0.00, V-J = 0.45, V-H = 0.56, V-K = 0.67 where as HR 8342 shows the following colors: B-V = -0.008 V-J = 0.05 V-H =0.01 V-K =0.036 almost text book colors for a A0 V star without any color excesses. Thus one is hard put to understand the light variability seen by Ragoonatha Chary and other Madras Observers. On the other hand they are experienced observers whose estimations of magnitudes cannot easily be ignored. Thus the variability of 'U Cephei, var 5.' in Madras catalogues remains as a mystery and the credit of discovery of the variability to Ragoonatha Chary remains in doubt. It is surprising why the observations of this star were not pursued or commented on later either by Pogson or other assistants after the death of Ragoonatha Chary in February 1880.

## 5     CONCLUDING REMARKS

Ragoonatha Chary was passionate to the cause of improving Astronomy in India. In his talk given at Pacheappah's Hall, Madras, on 13 April 1874 his plea for native observatory reflects this

> In Europe, excluding Russia, there now exist fifty-four public and ten private observatories spread over an area of less than two million square miles. In India with a surface of one and half million miles we have but one, that one wholly supported by the State. I recommend no more than modest but thorough place of instruction and study should be founded where the theoretical knowledge can be united to actual practical work. ... Such places exist in hundreds in Europe, but nowhere is the need for them greater than in India. Not much money, a little zeal, a little steadfastness of purpose, wed these to a regard for science, and soon would the metropolis of southern India be graced with an Institution which would be an honor to the country.

It has been stated that the Observatories like the one at Madras, Bombay, Calcutta remained in effect "alien outposts of a foreign science" and are a kind of islands which solely served the British science (Ansari, 1985). Ragoonatha Chary was one person who strove constantly to spread the benefit of their existence with local public and not to let them remain as islands. Whenever a major celestial event occurred, e.g. total solar eclipses or transit of planets (Venus) over the Sun, he took the opportunity to publish pamphlets not only in English but in various local languages both to explain the phenomenon as well as how to improve the native astronomical methods and calculations with improved data. In his pamphlet on the Transit of Venus in 1874he states what prompted him to bring out the pamphlet. It ` is written principally for information of such of my countrymen as have not had the advantage of any regular course of scientific reading. ... Although the class of phenomena to which the Transit of Venus belongs is mentioned in Hindu treatise on Astronomy, especially of the Sidhanta Siromani, yet the Sidhantis or Hindu astronomers are really not familiar with the nature of this particular occurrence and cannot predict it with even a rough approach to accuracy, happening as it does at such strange and rare intervals'. He himself wrote the English version in one style – as a dialogue, as he had been accustomed to discuss astronomical facts and methods orally with Hindu professors. The essay was presented in that style (samvadam). He intentionally wrote the other language versions, Sanskrit, Canarese, Tamil, Telugu, Urdu (Persian), Malayalam, and Marathi in a different style "I have found convenient to vary the arrangement" as to cater to native public. "The cherished object" of his life was to bring out the monograph "upon Astronomy which should embody the corrections, equations and formulae established by European (modern) research together with what is proper to retain from our own works, and thus to construct a manual accessible to Hindu astronomers." Ill health took him away, on 5 February 1880, before he could complete his cherished object of life, but he remains as a symbol: - a nationalist who tried to evolve the traditional system towards a contemporary and progressive seeker of knowledge.

## 6     ACKNOWLEDGEMENTS

The authors would like to thank Dr. A.V. Raveendran for his advice and help. We appreciate the generosity of Ms Cherry Armstrong in making available a collection of photographs of N.R. Pogson, her great, great, grand-father and his family to us and further donating to IIA Archives. We also would like to thankfully acknowledge the help received from Tamilnadu Archives.

We finally would like to thank Department of Science and Technology (DST) Govt. of India for the financial assistance through the project SR/S2/HEP-26/06 .

**8       FIGURES**

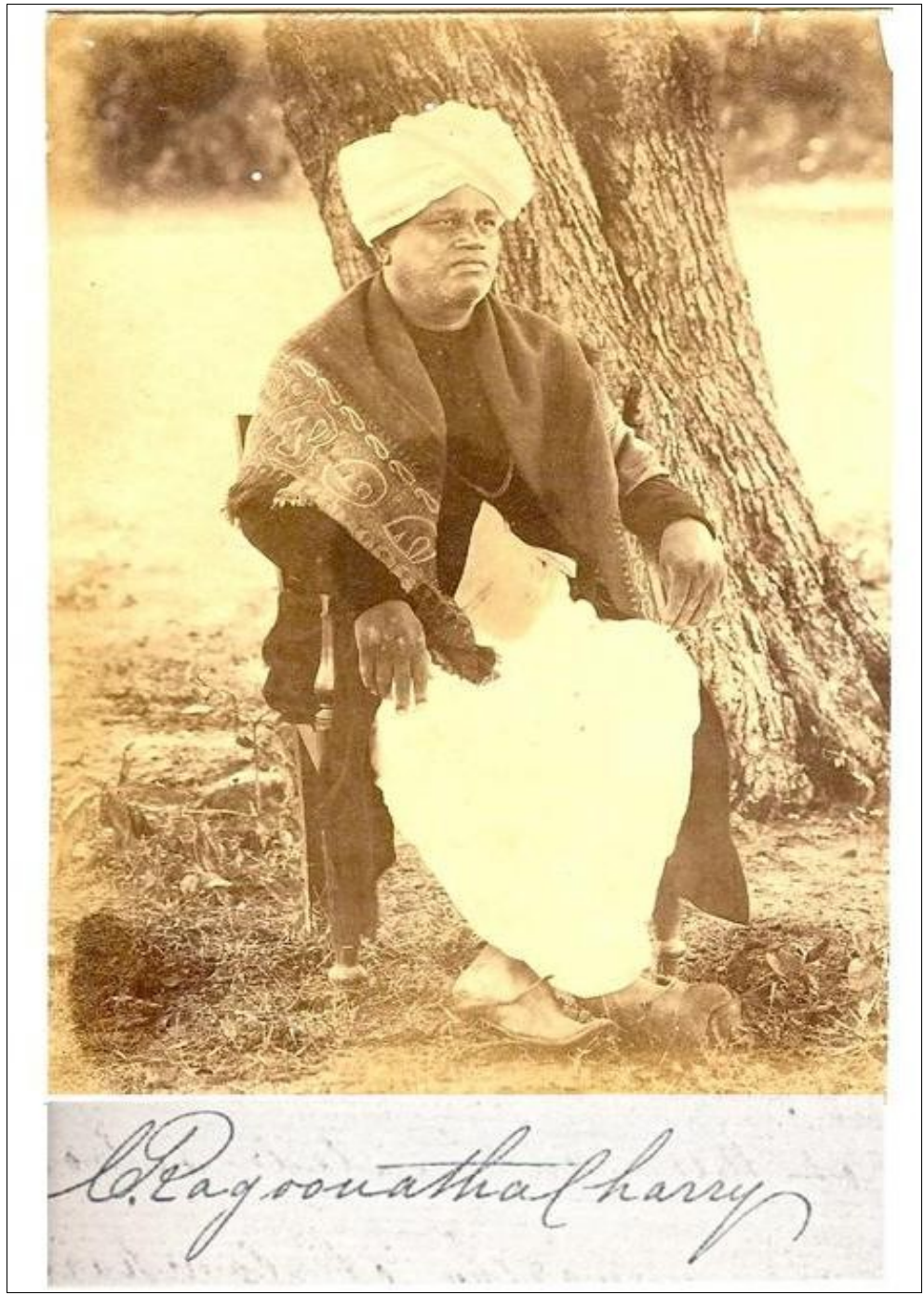

Figure 1. The picture of Ragoonatha Chary at Madras Observatory - from the collection of Ms. Cherry Armstrong, great, great, grand-daughter of N.R. Pogson.

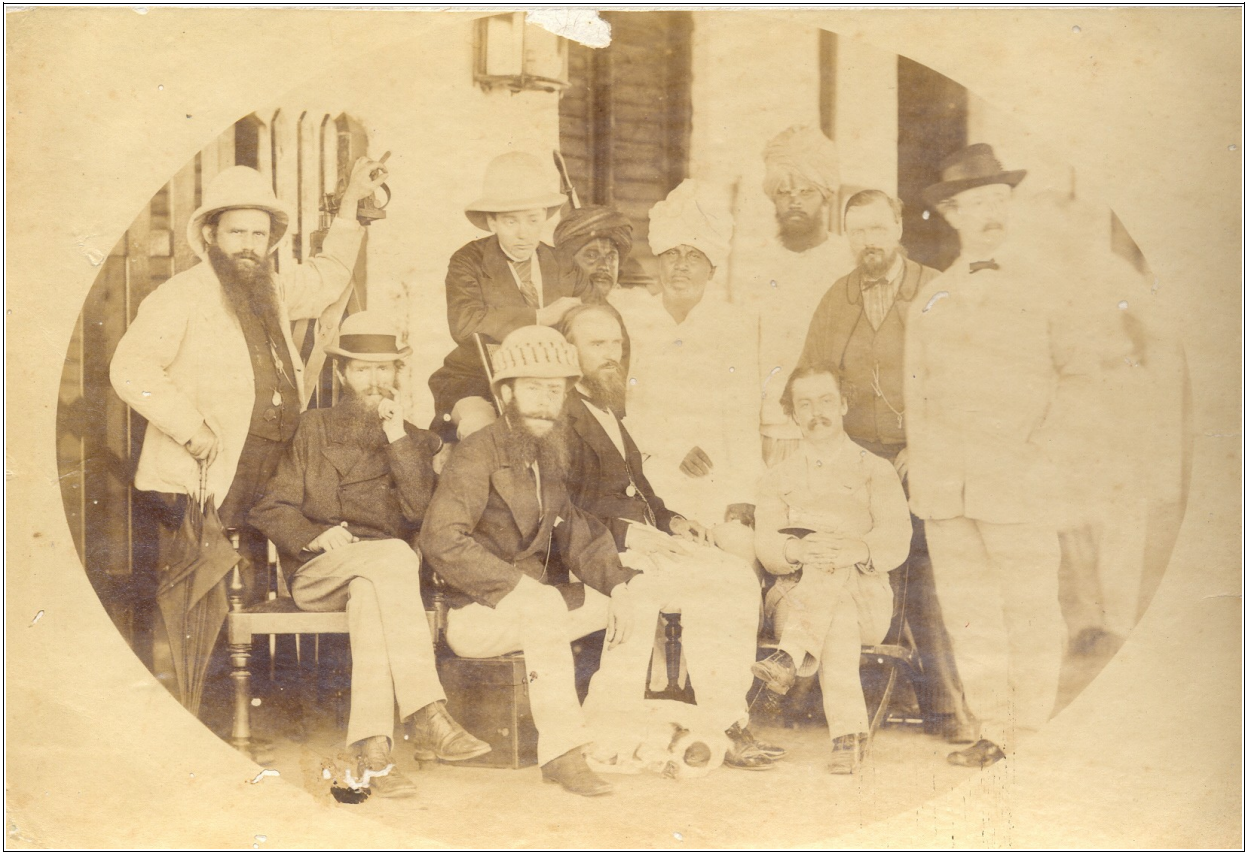

Figure 2. The picture of Ragoonatha Chary (next to N.R. Pogson, with a white turban) at Madras Observatory along with N.R. Pogson (centre, sitting with beard) and Norman Everard Pogson (second from left) possibly taken during 1871 around the time of total solar eclipse - from the collection of Ms. Cherry Armstrong, great, great, grand-daughter of N.R. Pogson. (We arrived at the identification of Ragoonatha Chary by process of elimination)

Separate Results of Madras Meridian Circle Observations in 1867.

| Number and Date. | Magnitude. | Mean Right Ascension 1867. h. m. s. | No. of Wires. | Mean Polar Distance 1867. ° ′ ″ | Observer. | Number and Date. | Magnitude. | Mean Right Ascension 1867. h. m. s. | No. of Wires. | Mean Polar Distance 1867. ° ′ ″ | Observer. |
|---|---|---|---|---|---|---|---|---|---|---|---|
| **160** | | *U Tauri* Var. 7. | | | | Jan. 15 | ... | 4 28 17·49 | ... | 73 45 39·8 | R |
| | | | | | | Oct. 15 | ... | 28 17·44 | ... | 45 39·7 | M |
| Jan. 17 | 9·9 | 4 14 4·25 | ... | 70 30 15·3 | R | 16 | ... | 28 17·51 | ... | 45 37·5 | R |
| **161** | | *Anon.* | | | | Dec. 9 | ... | 28 17·47 | ... | 45 40·4 | M |
| | | | | | | 12 | ... | 28 17·56 | ... | 45 40·3 | M |
| Nov. 16 | 10·0 | 4 15 20·86 | 5 | 129 7 25·1 | R | 21 | ... | 28 17·41 | ... | 45 39·9 | R |
| **162** | | *Anon.* | | | | **168** | | *Anon.* | | | |
| Nov. 14 | 8·0 | 4 16 49·23 | ... | 149 3 59·5 | M | Jan. 4 | 8·3 | 4 28 32·46 | 5 | 140 13 54·9 | M |
| | | | | | | Nov. 14 | 8·1 | 28 32·50 | ... | 13 58·0 | M |
| **163** | | 74 *Tauri* e | | | | | | | | | |
| Jan. 4 | ... | 4 20 51·28 | ... | 71 7 3·0 | M | **169** | | *R Reticuli* Var. 1. | | | |
| 8 | ... | 20 51·04 | ... | 7 4·2 | M | | | | | | |
| 9 | ... | 20 51·07 | ... | 7 3·9 | M | Jan. 16 | 10·0 | 4 32 10·56 | 6 | 153 18 22·0 | R |
| 10 | ... | 20 51·10 | ... | 7 3·8 | M | 17 | 9·8 | 32 10·21 | ... | 18 22·2 | R |
| 11 | ... | 20 51·20 | ... | 7 3·3 | M | 23 | 9·5 | 32 10·40 | ... | 18 21·2 | R |
| 12 | ... | 20 51·20 | ... | 7 3·0 | M | Feb. 1 | 8·6 | 32 10·31 | 4 | 18 21·4 | M |
| 14 | ... | 20 51·24 | ... | 7 3·3 | R | 15 | 7·7 | 32 10·49 | ... | 18 19·3 | M |
| Oct. 15 | ... | 20 50·97 | ... | 7 3·5 | M | Nov. 16 | 9·3 | 32 10·56 | 5 | 18 20·4 | R |
| Dec. 9 | ... | 20 51·19 | ... | 7 3·4 | M | Dec. 14 | 7·8 | 32 10·14 | ... | 18 20·0 | M |
| 16 | ... | 20 51·07 | ... | 7 3·1 | R | | | | | | |
| 19 | ... | 20 51·16 | ... | 7 2·2 | R | **170** | | *Lacaille* 1551. | | | |
| **164** | | *Lacaille* 1519. | | | | Jan. 24 | 6·0 | 4 32 11·35 | 5 | 153 5 52·9 | R |
| Jan. 7 | 7·0 | 4 25 37·48 | 5 | 153 5 43·2 | M | **171** | | *Anon.* | | | |
| **165** | | *Lacaille* 1520. | | | | Dec. 16 | 9·2 | 4 34 39·30 | ... | 153 26 30·7 | R |
| Jan. 10 | 7·7 | 4 26 44·46 | ... | 147 27 39·0 | M | 21 | 9·7 | 34 39·34 | 5 | 26 31·5 | R |
| **166** | | *Anon.* | | | | **172** | | 95 *Tauri.* | | | |
| Dec. 13 | 9·3 | 4 27 15·77 | 6 | 150 33 35·1 | M | Jan. 8 | 6·9 | 4 35 10·70 | 4 | 66 10 0·0 | M |
| **167** | | 87 *Tauri* α, Aldebaran. | | | | **173** | | *Lacaille* 1567. | | | |
| Jan. 3 | ... | 4 28 17·46 | ... | 73 45 39·9 | M | Jan. 15 | 5·0 | 4 35 18·10 | 5 | 152 20 28·5 | R |
| 5 | ... | 28 17·37 | ... | 45 39·6 | M | | | | | | |
| 8 | ... | 28 17·37 | ... | 45 41·1 | M | **174** | | *Lacaille* 1566. | | | |
| 9 | ... | 28 17·43 | 5 | 45 40·7 | M | | | | | | |
| 12 | ... | 28 17·43 | ... | 45 41·5 | M | Jan. 10 | 6·9 | 4 35 48·90 | ... | 143 28 6·2 | M |
| 14 | ... | 28 17·43 | ... | 45 39·1 | R | | | | | | |

Figure 3. The table in 'Separate Results of Madras Meridian Circle Observations in 1867'' showing the observations of R Reticuli

Separate Results of Madras Meridian Circle Observations in 1878.

| Number and Date. | Magnitude. | Mean Right Ascension 1878. h. m. s. | No. of Wires. | Mean Polar Distance 1878. ° ′ ″ | Observer. | Number and Date. | Magnitude. | Mean Right Ascension 1878. h. m. s. | No. of Wires. | Mean Polar Distance 1878. ° ′ ″ | Observer. |
|---|---|---|---|---|---|---|---|---|---|---|---|
| **873** | | μ Cygni—2nd. | | | | **881** | | 16 Pegasi. | | | |
| Oct. 1 | ... | 21 38 41·34 | ... | 61 48 28·1 | C.R | Oct. 2 | ... | 21 47 30·62 | ... | 64 38 53·1 | C.R |
| 19 | ... | 38 41·50 | 6 | 48 29·0 | C.R | 3 | ... | 47 30·66 | ... | 38 52·4 | C.R |
| 25 | ... | 38 41·49 | ... | 48 27·6 | C.R | | | | | | |
| **874** | | 9 Pegasi. | | | | **882** | | 30 Aquarii. | | | |
| Sep. 24 | 4·5 | 21 38 43·96 | ... | 73 12 31·2 | R | Sep. 21 | 5·0 | 21 56 51·49 | ... | 97 6 37·9 | R |
| 30 | 4·5 | 38 44·17 | 4 | 12 29·7 | R | 25 | 5·5 | 56 51·32 | ... | 6 39·7 | R |
| Oct. 21 | 5·0 | 38 43·94 | ... | 12 31·9 | C.R | 27 | 5·5 | 56 51·37 | ... | 6 38·9 | R |
| | | | | | | Oct. 21 | 5·7 | 56 51·27 | ... | 6 40·7 | C.R |
| **875** | | 10 Pegasi κ | | | | 24 | ... | 56 51·40 | ... | 6 38·9 | C.R |
| Sep. 23 | ... | 21 39 7·10 | ... | 64 54 54·2 | R | **883** | | 16 Cephei. | | | |
| 28 | 4·0 | 39 7·25 | ... | 54 54·6 | R | Oct. 1 | ... | 21 57 29·91 | 5 | 17 24 0·9 | C.R |
| Oct. 24 | 4·7 | 39 7·12 | ... | 54 54·0 | C.R | 8 | ... | 57 29·75 | ... | 24 2·3 | C.R |
| | | | | | | 22 | 5·0 | 57 29·94 | ... | 24 3·5 | C.R |
| **876** | | 11 Cephei. | | | | Nov. 8 | 5·0 | 57 30·28 | ... | 24 3·0 | M |
| Nov. 14 | 4·6 | 21 40 7·88 | ... | 19 14 59·6 | M | 9 | 5·2 | 57 30·34 | ... | 24 1·3 | M |
| **877** | | 10 Cephei ν | | | | **884** | | Anon. | | | |
| Sep. 21 | 4·5 | 21 41 55·75 | ... | 29 26 29·7 | R | Sep. 24 | 10·0 | 21 57 50·26 | ... | 92 31 10·4 | R |
| | | | | | | 28 | 10·4 | 57 50·33 | 4 | 31 7·6 | R |
| **878** | | 81 Cygni π² | | | | Oct. 28 | 9·9 | 57 50·45 | ... | 31 10·2 | C.R |
| Oct. 4 | ... | 21 42 16·98 | ... | 41 15 16·1 | C.R | **885** | | 34 Aquarii α | | | |
| | | | | | | *Oct. 3 | ... | 21 59 30·94 | ... | 90 54 42·7 | C.R |
| **879** | | 14 Pegasi. | | | | 29 | ... | 59 31·04 | ... | 54 43·9 | C.R |
| Sep. 27 | 5·0 | 21 44 26·71 | ... | 60 28 35·0 | R | Nov. 2 | ... | 59 30·92 | ... | 54 43·4 | C.R |
| 30 | 5·0 | 44 26·77 | ... | 28 34·8 | R | 6 | ... | 59 30·99 | ... | 54 41·7 | M |
| Oct. 21 | 5·0 | 44 26·78 | ... | 28 36·4 | C.R | 21 | ... | 59 31·05 | ... | 54 43·1 | M |
| 23 | 5·0 | 44 26·70 | ... | 28 34·3 | C.R | **886** | | 18 Cephei. | | | |
| | | | | | | Oct. 17 | 5·5 | 22 0 13·74 | ... | 27 28 24·1 | C.R |
| **880** | | υ Cephei, var 5. | | | | Nov. 11 | 5·4 | 0 13·97 | ... | 28 24·0 | M |
| Sep. 18 | 5·0 | 21 44 51·40 | ... | 20 24 51·9 | R | 14 | 5·5 | 0 13·75 | ... | 32 23·7 | M |
| 19 | 5·0 | 44 51·32 | ... | 24 51·5 | R | **887** | | 24 Pegasi ι | | | |
| Oct. 17 | 8·2 | 44 51·80 | 6 | 24 52·7 | C.R | | | | | | |
| 22 | 9·0 | 44 51·44 | ... | 24 52·9 | C.R | Oct. 24 | 4·0 | 22 1 19·87 | 5 | 65 15 0·5 | C.R |
| Nov. 6 | 7·8 | 44 51·88 | ... | 24 51·8 | M | | | | | | |

88

Figure 4. The table in 'Separate Results of Madras Meridian Circle Observations in 1878' showing the observations of U Cephei var.5

Separate Results of Madras Meridian Circle Observations in 1880.

| Number and Date. | Magnitude. | Mean Right Ascension 1880. h. m. s. | No. of Wires. | Mean Polar Distance 1880. ° ′ ″ | Observer. | Number and Date. | Magnitude. | Mean Right Ascension 1880. h. m. s. | No. of Wires. | Mean Polar Distance 1880. ° ′ ″ | Observer. |
|---|---|---|---|---|---|---|---|---|---|---|---|
| Oct. 5 | ... | 21 25 14.44 | ... | 96 5 53.8 | M | Oct. 14 | ... | 21 47 36.15 | ... | 64 38 16.4 | M |
| 6 | ... | 25 14.48 | ... | 5 52.2 | M | 21 | ... | 47 36.13 | ... | 38 17.0 | M |
| 9 | ... | 25 14.44 | ... | 5 52.0 | M | 25 | ... | 47 36.08 | ... | 38 15.4 | M |
| 20 | ... | 25 14.33 | ... | 5 52.1 | M | 26 | ... | 47 36.25 | ... | 38 18.1 | M |
| 26 | ... | 25 14.42 | ... | 5 52.7 | M | Nov. 5 | ... | 47 36.07 | ... | 38 19.0 | R |
| Nov. 1 | ... | 25 14.34 | ... | 5 55.1 | R | | | | | | |
| **542** | | **8 Pegasi ε** | | | | **547** | | **W. B. E. XXI. 1334.** | | | |
| Oct. 2 | ... | 21 38 17.49 | ... | 80 40 26.8 | M | Oct. 13 | 8.0 | 21 59 12.08 | ... | 98 16 38.8 | M |
| 21 | ... | 38 17.52 | ... | 40 25.1 | M | 16 | ... | 59 11.98 | ... | 16 38.3 | M |
| 25 | ... | 38 17.54 | ... | 40 27.0 | M | 25 | ... | 59 11.85 | ... | 16 39.5 | M |
| 26 | ... | 38 17.42 | ... | 40 28.5 | M | 26 | ... | 59 12.04 | ... | 16 40.3 | M |
| Nov. 1 | ... | 38 17.58 | ... | 40 27.1 | R | 29 | 8.0 | 59 11.98 | ... | 16 39.1 | M |
| 2 | ... | 38 17.54 | ... | 40 26.8 | R | | | | | | |
| **543** | | **78 Draconis.** | | | | **548** | | **34 Aquarii a** | | | |
| Sep. 16 | ... | 21 41 35.97 | ... | 18 13 47.8 | R | Oct. 2 | ... | 21 59 37.11 | ... | 90 54 5.9 | M |
| 18 | ... | 41 36.00 | ... | 13 46.6 | R | 6 | ... | 59 37.07 | ... | 54 6.5 | M |
| 22 | ... | 41 35.97 | ... | 13 47.3 | R | 8 | ... | 59 37.12 | ... | 54 5.7 | M |
| 24 | ... | 41 36.01 | ... | 13 46.1 | R | Nov. 8 | ... | 59 37.06 | ... | 54 5.3 | R |
| 30 | ... | 41 35.81 | ... | 13 47.5 | R | | | | | | |
| Oct. 1 | ... | 41 36.13 | ... | 13 45.3 | M | **549** | | **17 Cephei ξ—2nd.** | | | |
| **544** | | **10 Cephei ν** | | | | Sep. 17 | ... | 22 0 19.00 | ... | 25 57 21.3 | R |
| Sep. 15 | ... | 21 41 59.00 | ... | 29 25 56.4 | R | 18 | ... | 0 18.87 | ... | 57 21.7 | R |
| 17 | ... | 41 59.05 | ... | 25 57.2 | R | 21 | ... | 0 18.79 | ... | 57 21.6 | R |
| 21 | ... | 41 58.96 | ... | 25 57.7 | R | 22 | ... | 0 18.97 | ... | 57 20.4 | R |
| 23 | ... | 41 59.12 | ... | 25 57.2 | R | 23 | ... | 0 19.21 | ... | 57 21.2 | R |
| **545** | | **U Cephei, Var. 5.** | | | | **550** | | **Anon.** | | | |
| Oct. 2 | ... | 21 44 53.65 | ... | 20 24 19.0 | M | Oct. 21 | 9.5 | 22 1 43.95 | ... | 98 31 57.0 | M |
| 4 | 6.5 | 44 53.86 | ... | 24 19.1 | M | 27 | 9.5 | 1 44.12 | ... | 31 56.3 | M |
| 5 | 6.8 | 44 53.69 | ... | 24 20.6 | M | 28 | 9.5 | 1 44.08 | ... | 31 56.8 | M |
| 6 | 7.0 | 44 53.84 | ... | 24 18.9 | M | 30 | 9.5 | 1 43.93 | ... | 31 58.3 | M |
| 7 | 7.0 | 44 53.60 | ... | 24 19.2 | M | Nov. 2 | 9.6 | 1 44.02 | 5 | 31 57.6 | R |
| **546** | | **16 Pegasi.** | | | | | | | | | |
| Sep. 17 | ... | 21 47 36.06 | ... | 64 38 18.6 | R | **551** | | **15 Piscis Australis.** | | | |
| 22 | ... | 47 36.10 | ... | 38 16.6 | R | | | | | | |
| 23 | ... | 47 36.16 | ... | 38 16.4 | R | Sep. 6 | 5.6 | 22 3 6.58 | ... | 123 8 14.0 | R |
| Oct. 8 | ... | 47 36.13 | ... | 38 18.6 | M | 15 | 5.6 | 3 6.80 | ... | 8 12.7 | R |

Figure 5. The table in 'Separate Results of Madras Meridian Circle Observations in 1880' showing the observations of U Cephei Var.5